# Experimental and Numerical Investigation of the Fracture Behavior of Particle Reinforced Alkali Activated Slag Mortars

Sumeru Nayak[1], Ahmet B. Kizilkanat[2], Narayanan Neithalath[3], Sumanta Das[4*]

[1] Graduate Student, Civil and Environmental Engineering, University of Rhode Island, Kingston, RI, USA

[2] Assistant Professor, Department of Civil Engineering, Yildiz Technical University, Istanbul, Turkey.

[3] Professor, School of Sustainable Engineering and the Built Environment, Arizona State University, Tempe, AZ, USA

[4*]Assistant Professor, Civil and Environmental Engineering, University of Rhode Island, Kingston, RI 02881, USA, Email: sumanta_das@uri.edu (corresponding author)

**ABSTRACT**

This paper presents fracture response of alkali-activated slag (AAS) mortars with up to 30% (by volume) of slag being replaced by waste iron powder which contains a significant fraction of elongated particles. The elongated iron particles act as micro-reinforcement and improve the crack resistance of AAS mortars by increasing the area of fracture process zone (FPZ). Increased area of FPZ signifies increased energy-dissipation which is reflected in the form of significant increase in the crack growth resistance as determined from R-curves. Fracture response of notched AAS mortar beams under three-point bending is simulated using extended finite element method (XFEM) to develop a tool for direct determination of fracture characteristics such as crack extension and fracture toughness in particulate-reinforced AAS mortars. Fracture response simulated using the XFEM based framework correlates well with experimental observations. The comprehensive fracture studies reported here provide an economical and sustainable means towards improving the ductility of AAS systems which are generally more brittle than their conventional ordinary portland cement counterparts.

## 1. INTRODUCTION

Alkali activated ground granulated blast furnace slag (GGBFS) has emerged as a sustainable alternative to ordinary portland cement (OPC)-based binders primarily because of the fact that production of ground granulated blast furnace slag (GGBFS) results in lower energy and carbon footprints than OPC production (Bernal et al. 2011; Collins and Sanjayan 2000; Wang et al. 1994; Wang and Scrivener 1995). Alkali silicates are the commonly used activating agent for slag. The main reaction product in these binders is calcium (alumino) silicate hydrate, C-(A)-S-H gel, similar to C-S-H gel in conventional portland cement-based binders (Pacheco-Torgal et al. 2008; Pan et al. 2018). The fresh and hardened properties of alkali activated slag (AAS) binders have been characterized in detail in various studies (Ding Yao et al. 2018; El-Wafa Mahmoud Abo and Fukuzawa Kimio 2018; Najimi Meysam et al. 2018; Radlińska Aleksandra et al. 2013; Shi et al. 2003; Wang et al. 1995). Although the flexural and compressive strengths of AAS binders have been reported in detail in several studies(Collins and Sanjayan 1999; Duran Atiş et al. 2009; Fernández-Jiménez et al. 1999; Talling and Brandstetr 1989), limited studies exist on the fracture response of such binders, primarily because they are known to be generally very brittle. One of the means to reduce the brittleness of such mixtures is the use of metallic reinforcement, which is usually incorporated in the form of fibers. In this study, we report the use of a waste metallic powder as an additive to AAS mortars to enhance the crack resistance of such systems. A well-established two parameter fracture model (TPFM) (Das et al. 2014a, 2015a) coupled with digital image correlation (DIC) (Dakhane et al. 2016) is used to quantify the beneficial effects of metallic waste power on the fracture performance of mortars.

The waste iron powder used in this study is a byproduct from the electric arc furnace (EAF) manufacturing process of steel and shot-blasting of structural steel (Machado et al. 2006; Sofilić et al. 2004). This waste dust is primarily landfilled since recycling of iron from the dust is not economically feasible (Das et al. 2014b). Several million tons of such waste iron dust is being landfilled all over the world at a great cost. Hence, potential use of this waste iron dust in AAS mortars as replacement of slag would reduce the demand of slag in these binders. Since ground granulated blast furnace slag (GGBFS) is a marketed commodity with no excess supply in the United States, reducing slag use is an economical means to produce durable concrete. However, it has been reported that alkali activated slag mixtures are much more brittle than conventional portland cement mixtures (Thomas and Peethamparan 2015). Here, we employ the use of metallic particulate reinforcement than the commonly used metallic fiber reinforcement to enhance the ductility of such systems. Moreover, the metallic particulate used here is a waste material from steel shot blasting, thereby providing the composite with sustainability benefits. We evaluate the fracture behavior of the waste iron powder-incorporated AAS mortars using compliance-based resistance curves (R-curves) (Das et al. 2014a, 2015a; c, 2016; Das and Neithalath 2016) determined from three-point bending tests.

Fundamental characterization of fracture response in quasi-brittle materials also requires direct observation of fracture process zone (FPZ) which is denoted by the zone of strain localization near the tip of the advancing crack. FPZ has been geometrically quantified using microscopy (Hadjab.S et al. 2007; Nemati 2006), photography (Bhargava and Rehnström 1975) or a non-contact speckle-tracking method called digital image correlation(DIC) (Das et al. 2014a, 2015b; Skarżyński et al. 2013; Yates et al. 2010). In DIC, the surface displacements and strain maps are obtained by correlating the images and the direct measurements of the crack extensions/FPZ are quantified from the displacement/strain maps. In order to shed more light into the fundamental difference in crack propagation and strain localization behavior imparted by iron particulates in AAS mortars, direct measurements of crack extension and fracture process zone are made using digital image correlation (DIC) technique. Furthermore, correlations are drawn between the FPZ characteristics and the crack growth resistance. We have also simulated the fracture response of notched AAS mortar beams under three-point bending using the extended finite element method (XFEM) (Belytschko and Black 1999a; Benson et al. 2010) so as to predict crack extension and fracture toughness in particulate-reinforced AAS mortars. The paper thus combines experimental characterization of fracture response and the process zone along with numerical simulation to provide detailed insights into the fracture response of these systems.

## 2. EXPERIMENTAL PROGRAM

### 2.1 Materials and mixture proportions

The materials used in the experiments are commercially available Type I/II OPC complying with ASTM C 150 and ground granulated blast furnace slag (GGBFS) Type 100 complying with ASTM C 989. Table 1 shows the chemical composition of OPC and GGBFS along with their median particle sizes. The iron powder used in the experiments is a waste powder obtained from shot-blasting of structural steel sections, which is expensive and difficult to dispose. It has an iron content of 88% and oxygen content of 10% which is due to the atmospheric oxidation of iron apart from trace elements like copper, manganese and calcium. Four GGBFS mortar mixtures were prepared where the sand had an average particle size of 0.6 mm, and varying amounts of iron powder waste (0, 10, 20, 30%) by volume replaced slag. The volume of sand in the mortar was kept at 50%. Potassium silicate activator solutions were proportioned based on the $K_2O$-to-slag (binder) ratio (n) and the silica modulus of the activator (molar $SiO_2$-to-$K_2O$ ratio) ($M_s$). A $K_2O$-to-slag (binder) ratio (n) of 0.05 was adopted here based on previous studies (Dakhane et al. 2016).The potassium silicate solution has a solids content of 44% and a molar $M_s$ of 3.29. KOH was added to bring the activator $M_s$ down to 1.5 which has been shown to provide strength equivalent to hardened cement paste (Chithiraputhiran and Neithalath 2013; Ravikumar and Neithalath 2012). A constant liquid-to-slag ratio of 0.50 by mass was maintained. For compressive strength tests, 50 mm mortar cubes were used. For flexural strengths, prismatic mortar beams of size 250 mm x 50 mm x 50 mm were used. For fracture tests, 330 mm long notched beams with a cross section of 25 mm x 75 mm were used. Four replicates were used in all the mechanical property tests. Each sample was demolded after 24 hours and exposed to a moist environment with relative humidity greater than 98% and the curing temperature was maintained at 23$\pm$2°C. The samples used for microstructural analysis by mercury intrusion porosimetry and scanning electron microscopy were kept sealed in containers.

Table 1: Composition and median size of constituent binders

| Material | Oxide composition (% by mass) | | | | | | | |
|---|---|---|---|---|---|---|---|---|
| | CaO | $SiO_2$ | $Al_2O_3$ | MgO | $SO_2$ | $K_2O$ | $Fe_2O_3$ | $d_{50}$ (µm) |
| OPC | 63 | 21 | 3.6 | 2.6 | 3.9 | 0.82 | 3 | 8.8 |
| GGBFS | 37.1 | 41.8 | 3.6 | 12.59 | 3 | 0.46 | 0.3 | 7.58 |

**2.2 Mercury Intrusion Porosimetry (MIP)**

The pore structure of the mortars was evaluated using mercury intrusion porosimetry (MIP). The low pressure run includes gas evacuation and exposing the sample to 345 kPa whereas the high pressure run exposes the sample to 414 MPa. The Washburn equation (Ghafari et al. 2014; Mohamed Elrahman and Hillemeier 2014; Moon et al. 2006; Sánchez-Fajardo et al. 2014) was used to determine the pore diameter. The surface tension of mercury was taken as 0.485 N/m while the contact angle between mercury and the pore wall was considered 130°. Although the pore-structure information obtained using MIP cannot be used for exact quantitative measurements (Diamond 2000), it is suitable for qualitative comparisons (Das et al. 2014a).

**2.3 Scanning electron microscopy for microstructural evaluation**

The sample preparation for the microstructural analysis involved cutting the sample into a cube of 2 mm edge length with a diamond saw followed by ultrasonic cleaning and alcohol rinsing to remove debris. The sample was then impregnated with epoxy by vacuum impregnation technique and cured overnight. Suitable grinding and polishing techniques were adopted to achieve a planar surface suitable for microscopic analysis. SiC discs were used to polish using successively finer abrasives. Finally the sample was polished with 0.04μm colloidal silica suspension and subjected to backscattered mode imaging in a Philips XL30 field emission environmental scanning electron microscope (FESEM).

## 2.4 Three Point Bending Test

The experimental setup for three-point-bending tests is shown in Figure 1. Notched mortar beams of size 330 mm (effective span 305 mm) x 25 mm (width) x 75 mm (depth) were used. The notch depth was chosen as a quarter of the total depth as shown in Figure 1. The test was performed under crack mouth opening displacement (CMOD)-controlled mode, with the CMOD monitored using a clip gauge . The test was carried out under load-controlled mode till a load of 100 N, beyond which CMOD-controlled mode was initiated. CMOD-controlled stage was terminated at a CMOD value of 0.38 mm following which unloading was initiated using load-controlled mode. The unloading rate was 556N/m, and was terminated at a load of 50 N. Next, CMOD-controlled mode was again implemented in the reloading cycle followed by load-controlled unloading cycles until a CMOD of 0.18 mm is reached.  The loading-unloading cycles, thus implemented in the load-CMOD response, are used in a compliance-based technique to obtain crack growth resistance values which is explained in detail later.

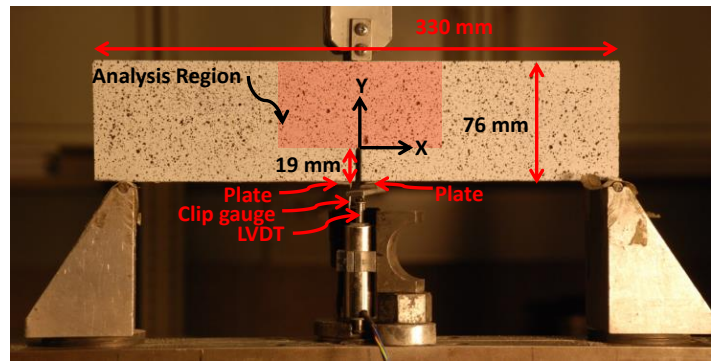

Figure 1: Experimental setup for closed loop CMOD-controlled three-point bending test

## 2.5 Digital Image Correlation (DIC)

Digital image correlation (DIC) is a non-contact optical method that tracks speckles to measure displacements on the surface of the specimen. A correlation is drawn between images of undeformed and deformed states to calculate displacement fields (Chen et al. 2010; Ghorbani et al. 2014; Hjelmstad 2007; Krottenthaler et al. 2013; Rossol et al. 2013; Sutton et al. 2009; Yates et al. 2010; Yuan et al. 2014). In this study, the beam surfaces were first painted white, and random black speckles were sprayed on them.  A CCD camera was used to capture images every 5 seconds during the CMOD-controlled three-point-bending test using VIC-snap software from Correlated Solutions [17,18]. Once the images were obtained, commercially available software VIC-2D was used to obtain the displacement and strain fields (Das et al. 2014a). The displacement/strain maps are used here to characterize the fracture process zone (FPZ) and crack propagation behavior.

## 3. RESULTS AND DISCUSSIONS

### 3.1 Pore- and Microstructure

A microstructural analysis of the polished samples after 28 days yields distribution of various component phases in the alkali-activated slag paste. Figure 2(a) shows a back scattered (BSE) image of hardened alkali-activated slag paste with 30% iron powder by volume. The bright (dense) elongated iron particles are clearly visible in the micrograph. The elongated shape of iron particles likely contributes towards improvement of fracture response through mechanisms such as crack bridging and crack deflection which are explored in detail in this paper.

Figure 2(b) shows the total porosity and average pore diameters, extracted using MIP, for the control AAS mortar as well as mortars containing various dosage of iron powder after 28 days. The total porosity and average pore diameters are not found to vary significantly with iron powder content. The similar liquid-to-slag ratio in all mixtures along with the fact that iron powder is an inert filler in the highly alkaline environment contributes to this response.

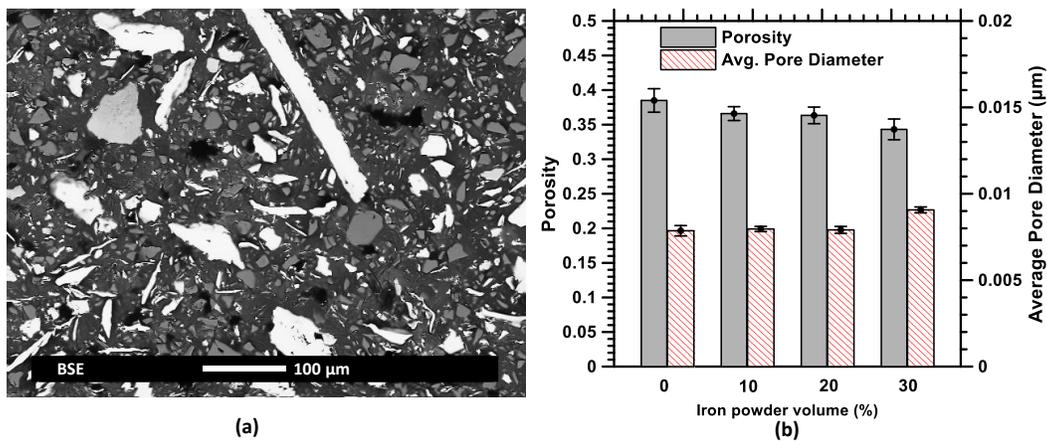

Figure 2: (a) BSE micrograph showing distribution of iron particulates in alkali-activated slag paste containing 30% iron powder by volume; (b) porosity and average pore diameter of the pastes with varying volume fraction of iron powder.

### 3.2 Compressive and Flexural Strengths

Figure 3 shows the compressive and flexural strengths of mortars after 28 days of curing. The compressive strengths are similar for all the mortars irrespective of iron powder content which can be attributed to the similar pore-structure features for these mortars as shown in Figure 2(b). However, the flexural strength shows an improvement with the increasing iron powder content, attributable to the elongated shape of iron particles as observed in Figure 2(a). These elongated iron particles act as micro-reinforcement to the matrix. Overall the strength results indicate that the waste iron powder can be safely incorporated to AAS mortar systems without compromising mechanical properties.

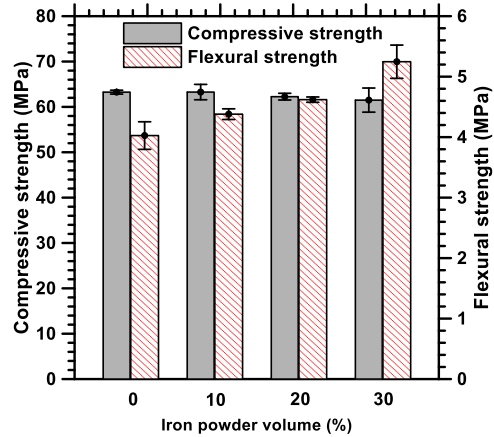

Figure 3: Compressive and flexural strengths of mortars with varying volume fractions of iron powder

### 3.3 Effect of Particulate-reinforcement on Fracture Behavior

While the previous section reported the strength and pore-structure of metallic particulate-reinforced AAS mortars, this section focusses on the influence of particulate reinforcement on the fracture response of AAS mortars.

#### *3.3.1 Load-CMOD Responses*

The load-CMOD responses of control and iron particle-incorporated AAS mortars are shown in Figure 4. Multiple loading-unloading cycles are implemented in the load-CMOD response in the post-peak regime in order to calculate unloading compliances corresponding to each cycle which are used to calculate compliance-based resistance curves as explained later in this paper. The peak load increases with increase in the amount of iron powder in the binder due to the micro-reinforcing effect of the elongated iron particulates. The area under the load-CMOD curve is a measure of the material toughness. An increase in the area under load-CMOD curve with increasing iron content indicates improved toughness of the material.

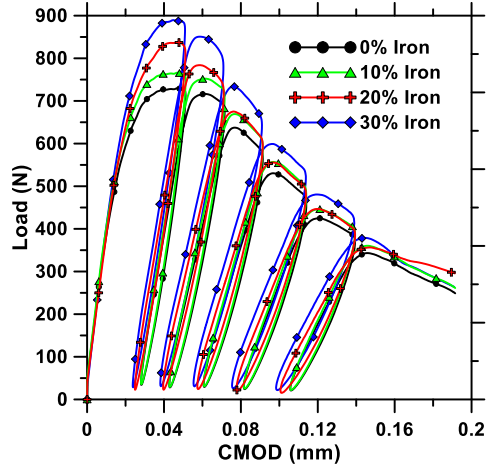

Figure 4: Load CMOD for AAS mortars with varying volume fractions of iron powder

### 3.3.2 Crack Growth Resistance

The fracture response of particulate-reinforced AAS mortars is quantified here using compliance-based crack growth resistance curves. The basic assumption behind the development of the compliance-based R-curves is that the compliance increases as the crack propagates. The total crack growth resistance ($G_R$) contains an elastic part that is calculated from the elastic compliance while the inelastic part is based on the inelastic CMOD. The crack growth resistance can be expressed as (Das et al. 2014a; Mobasher 2011; Wecharatana and Shah 1983):

$$G_R = G_{elastic} + G_{inelastic} = \frac{P^2}{2t}\frac{\partial C}{\partial a} + \frac{P}{2t}\frac{\partial (CMOD_{inelastic})}{\partial a} \qquad [1]$$

Where C is the unloading compliance, t is the thickness of specimen, P is the load and a is the crack length. The crack extension values are obtained from unloading compliances as explained in (Das et al. 2014a). The relationships between unloading compliances or the inelastic CMOD values and crack extensions are differentiated with respect to crack extension to obtain the rate terms of Equation 1 as explained in (Das et al. 2014a).

Figure 5 shows $G_R$ as a function of crack extension for all the mortars. The crack growth resistance increases with crack extension culminating in a relatively constant portion signifying that the fracture process zone has almost fully developed. It can be noticed from this figure that the crack growth resistance increases with increase in iron particulate content which is attributed to the crack bridging/deflection effect of the elongated iron particles as shown in the micrograph in Figure 2(a). The crack growth resistance is enhanced by about 50% through the addition of 30% iron powder in the mixture. It can also be observed from Figure 5 that the values of crack extension at which the plateau is reached decreases with increasing iron powder content which can be explained by the crack deflection effect of elongated iron particles. Such elongated iron particulates increase the tortuosity of the micro-cracks in the FPZ and dissipate a significant amount of energy resulting in achievement of a steady-state crack propagation at a lower value of crack extension in the presence of iron powder.

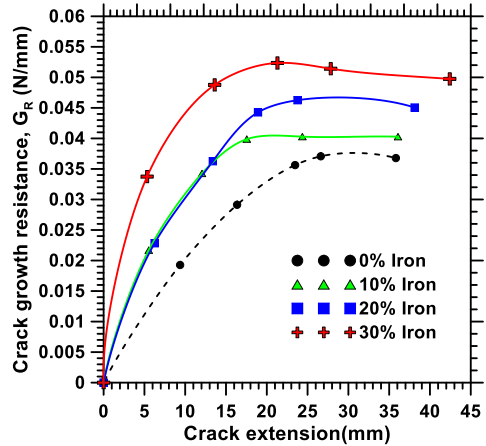

Figure 5: Crack growth resistance as a function of iron powder volume fraction

### 3.3.3 Evaluation of Fracture Process Zone (FPZ)

While the previous section evaluated the fracture response of the mortars through indirect quantifications, this section presents direct measurements of FPZ to shed more light on the influence of iron powder on the fracture response of AAS mortars. Direct characterization of the fracture process zone features including length, width and area is essential to elucidate the fundamental differences imparted by the iron particulates in terms of the crack extension and strain localization response. The geometric properties of FPZ are determined by DIC as explained in detail elsewhere (Das et al. 2014a, 2015a, 2016).

Figure 6 shows the Lagrangian strain fields for control AAS mortar as well as AAS mortars with varying iron powder contents at 95% of the peak load in the post-peak region. It can be seen that the length of FPZ increases with increase in iron powder content whereas the width of FPZ almost remains unchanged. At a location 4 mm above the tip of the notch, the quantified measurements of the FPZ at 95% of peak load in post-peak regime are shown in Figure 7, as a function of the iron powder volume fraction. The length of FPZ increases significantly with increasing iron powder content, indicating energy dissipation by various mechanisms such as micro-cracking, crack arresting and crack deflection due to the presence of elongated metallic particles. On the contrary, the width of the FPZ remains fairly constant irrespective of the iron powder content due to the fact that the inelastic component of crack growth resistance is not expected to change with inclusion of stiff particulates for the volume fractions considered here . Figure 7(b) shows the area of FPZ as a function of the iron powder content. The area of FPZ increases significantly with increasing iron powder content which is reflected in the form of overall increase in crack growth resistance shown in Figure 5.   To shed more light, the forthcoming section correlates the directly measured FPZ features to the crack growth resistances, calculated indirectly using compliance-based approach.

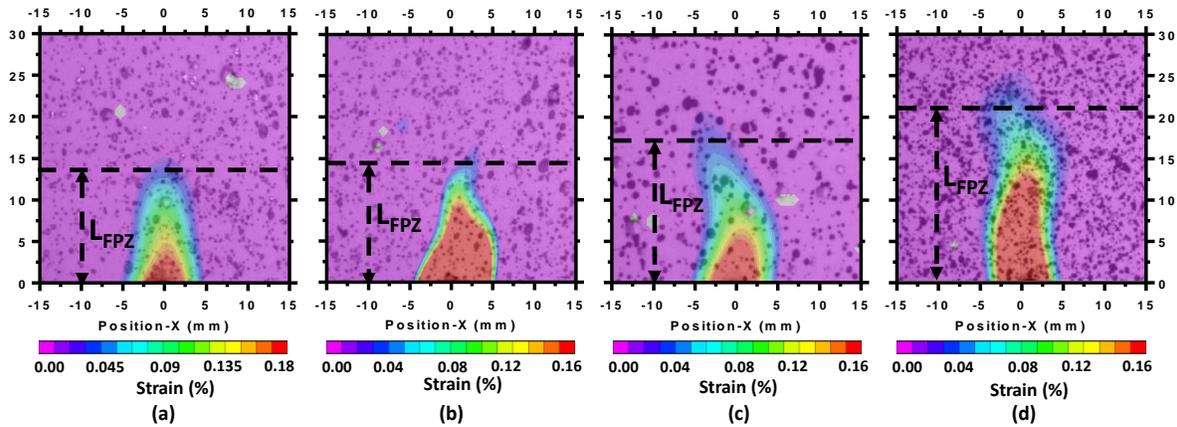

Figure 6: Lagrangian strain fields at 95% of the peak load in post-peak regime for: (a) control AAS mortar, and AAS mortars with: (b) 10%, (c)20% and (d) 30% waste iron powder as slag-replacement.

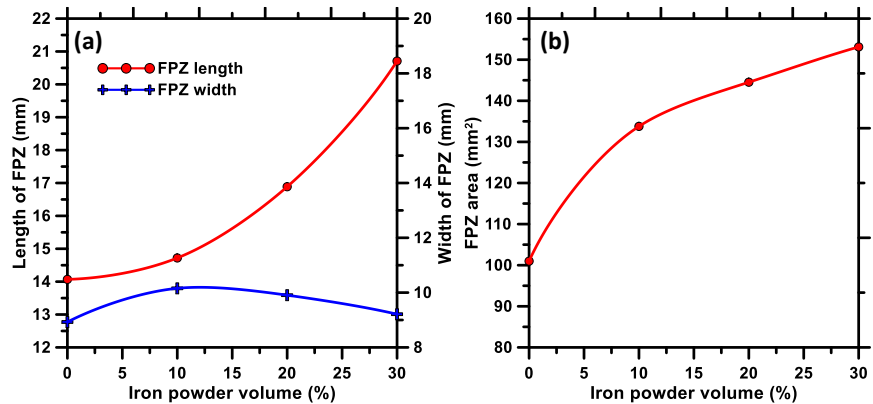

Figure 7: Relation between volume fraction of iron powder and: (a) FPZ geometry (b) FPZ area

### 3.3.4 Correlation between Crack Growth Resistance and FPZ

While it was mentioned in the previous section that the FPZ characteristics influence the energy dissipation in these mortars, this section quantifies such relationships. The FPZ characteristics are plotted against the crack growth resistances in order to draw a correlation between the two.  The FPZ parameters correspond to $0.95P_{max}$ in the post-peak regime for the control AAS mortar as well as mortars with varying iron powder contents. Figure 8(a) shows that the crack growth resistance increases with increase in the FPZ area, attributable to the increased area of energy dissipation with incorporation of iron powder. Figure 8(b) depicts the correlation between FPZ length and elastic component of crack growth resistance. As the volume fraction of iron powder increases in AAS mortar, the elastic component of the crack growth resistance also increases linearly with increase in the FPZ length. Incorporation of iron powder in AAS mortars facilitates microstructural strengthening and toughening of the mortar matrix through various mechanisms such as micro-cracking, crack-arresting and crack-deflection in the direction of the crack-driving force as described earlier. This microstructural strengthening and toughening mechanisms result in significant increase in the length of FPZ and consequently the elastic component of the crack growth resistance increases with incorporation of iron powder. Thus, this study has shown that metallic waste iron powder can be incorporated in AAS mortars as partial replacement of slag without compromising the strength, yet providing significant enhancement in the fracture response.

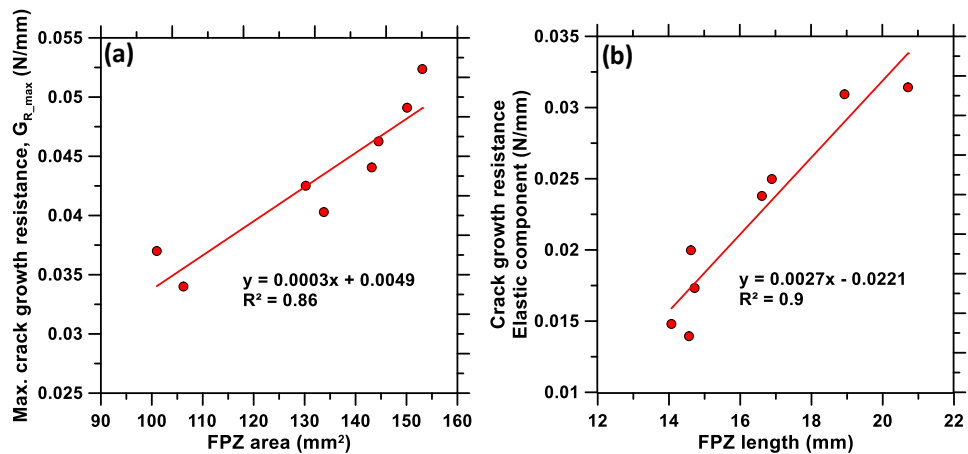

Figure 8: Relationships between: (a) FPZ area and crack growth resistance; (b) FPZ length and elastic components of crack growth resistance

### *3.3.5 Numerical simulation towards predicting the fracture behavior of mortars*

While the pervious section elucidated experimental evaluation of fracture behavior of particulate-reinforced AAS, this section is aimed towards prediction of fracture behavior of such systems using extended finite element method (XFEM). Crack-growth analysis using XFEM eliminates the need for re-meshing and thus provides an efficient solution(Belytschko and Black 1999a; Benson et al. 2010; Fleming et al. 1998; Moës et al. 1999; Zi and Belytschko 2003). Brief description of the formulations related to damage model and the prediction results are shown in the forthcoming sub-sections as follows:

### 3.3.5.1 XFEM for damage prediction in quasi-brittle materials:

Extended finite element method (XFEM) is a versatile tool for the analysis of problems characterized by discontinuities, singularities, localized deformations and complex geometries. In XFEM, local enrichment functions are incorporated in the FE approximation to model the crack (discontinuities) in an efficient manner. The enriched functions have additional degrees of freedom and simulate path-independent crack initiation and propagation based on the damage criteria provided. The approximation of displacement vector function with the partition of unity enrichment is given as (Belytschko and Black 1999b; Motamedi et al. 2013):

$$u = \sum_{I=1}^{N} N_I(x)[u_I + H(x)a_I + \sum_{\alpha=1}^{4} F_\alpha(x)b_I^\alpha] \quad (2)$$

where, $N_I(x)$ are the conventional nodal shape functions at node I, $u_I$ is the nodal displacement vector associated with the continuous part of the finite element solution; $a_I$ is the nodal enriched degree of freedom vector, $H(x)$ is the Heaviside function; $b_I^\alpha$ is the product of nodal enriched degree of freedom vector, and $F_\alpha(x)$ are the associated elastic asymptotic crack-tip functions. $H(x)$ is given as:

$$H(x) = 1 \text{ if } (x - x^*).n \geq 0 \quad (3)$$

If Equation 3 is not satisfied, it is equal to -1. Here, $x$ is a sample Gauss point, $x^*$ is the point on the crack closest to $x$, and $n$ is the unit outward normal to the crack at $x^*$. The asymptotic crack tip functions $F_\alpha$ are given as (Belytschko and Black 1999b):

$$F_\alpha(x) = [\sqrt{r}Sin\frac{\theta}{2}, \sqrt{r}Cos\frac{\theta}{2}, \sqrt{r}Sin\theta.Sin\frac{\theta}{2}, \sqrt{r}Sin\theta.Cos\frac{\theta}{2}] \quad (4)$$

where the crack tip is at the origin of the polar coordinate system and θ=0 is the tangent to the crack tip. The XFEM damage model requires appropriate damage initiation criteria. The maximum principal stress criterion is adopted in this study and the crack is considered to be initiated if the maximum principal stress exceeds the tensile strength of the mortar. A bilinear traction-separation law *(Bažant 2002; Roesler et al. 2007a)* is used for damage propagation, as shown in Equations 5a and 5b. The fracture energy ($G_F$) is the area under the entire traction-separation curve, given as:

$$G_F = \int_0^\infty f(w)dw \quad (5a)$$

$$f(w) = \left\{ f_t - (f_t - f_1)\frac{w}{w_1} \text{ for } w \leq w_1 \text{ and } \left| f_1 - f_1 \frac{(w-w_1)}{(w_c-w_1)} \text{ for } w_1 > w \right. \right\} \quad (5b)$$

where $f_t$ is the tensile strength of the material, $w_c$ is the critical crack tip opening displacement, and $f_1$ and $w_1$ are the stress and opening displacement corresponding to the kink in the bilinear traction-separation curve.

The numerical simulation framework incorporates the Concrete Damage-Plasticity (CDP) model *(Jankowiak and Lodygowski 2005)* beyond the linear elastic regime for fracture simulation. CDP is a material model based on a combination of damage and plasticity theory. Plasticity theory is used to describe both the compressive and tensile response of concrete, while the damage theory is used for the cyclic and unloading characteristics. In this model, isotropic damage is represented as:

$$\sigma = (1-d)D_0^{el} : (\varepsilon - \varepsilon^{pl}) = D^{el} : (\varepsilon - \varepsilon^{pl}) \tag{6}$$

where σ is the Cauchy stress tensor, d is the scalar stiffness degradation variable, ε is the strain tensor, $\varepsilon^{pl}$ is the plastic strain, $D_0^{el}$ is the initial elastic stiffness of the material, and $D^{el}$ is the degraded elastic stiffness tensor. The effective stress tensor $\overline{\sigma}$ is given as:

$$\overline{\sigma} = D_0^{el} : (\varepsilon - \varepsilon^{pl}) \tag{7}$$

Damage states in tension and compression are characterized independently by two hardening variables which are the equivalent plastic strains in compression and tension respectively. The plastic flow is given as *(Jankowiak and Lodygowski 2005)*:

$$\dot{\varepsilon}^{pl} = \dot{\lambda} \frac{\partial G(\overline{\sigma})}{d\overline{\sigma}} \tag{8}$$

where, the flow potential, G is given using a Drucker-Prager hyperbolic function as:

$$G = \sqrt{(f_c - m.f_t.\tan\beta)^2 + \overline{q}^2} - \overline{p}.\tan\beta - \sigma \tag{9}$$

Here, $f_t$ is the tensile strength and $f_c$ is the compressive strength, $\beta$ is the dilation angle and m is the eccentricity of the plastic potential surface, $\overline{p}$ is the effective hydrostatic stress and $\overline{q}$ is the Mises equivalent effective stress. The CDP model uses a yield condition based on loading function:

$$F = \frac{1}{1-\alpha}(\overline{q} - 3.\alpha.\overline{p} + \theta(\varepsilon^{pl})\langle\overline{\sigma}_{\max}\rangle - \gamma\langle-\overline{\sigma}_{\max}\rangle) - \overline{\sigma}_c(\varepsilon^{pl}) \tag{10}$$

where the function $\theta(\varepsilon^{pl})$ is given as:

$$\theta(\varepsilon^{pl}) = \frac{\overline{\sigma}_c(\varepsilon_c^{pl})}{\overline{\sigma}_t(\varepsilon_t^{pl})}(1-\alpha)-(1+\alpha) \qquad (11)$$

The parameter $\alpha$, which is based on the ratio of biaxial compressive strength ($f_{b0}$) to uniaxial compressive strength ($f_c$), is defined as:

$$\alpha = \frac{(f_{b0}/f_c)-1}{2(f_{b0}/f_c)-1} \qquad (12)$$

### 3.3.5.2 Simulation of three point bending tests of notched mortar beams:

The notched beam used for fracture experiments, explained in earlier sections, is modeled here using ABAQUS™ as shown in Figure 9.

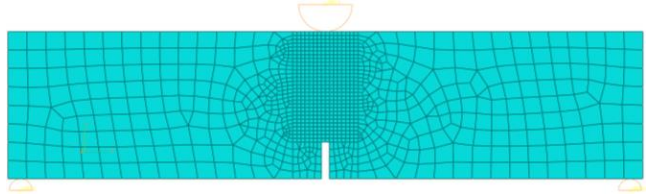

Figure 9: Assembly of a meshed notched beam in ABAQUS

The two pin supports have been modelled as analytical rigid bodies. The analytical rigid punch has been placed centrally with the displacement being applied on it using a reference point. Implementation of traction-separation law for propagation of cracks requires fracture energy and tensile strength of mortars. Here, fracture energies for all the mortars are calculated from the load-CMOD responses and the tensile strength values are obtained through a moment-curvature-based inverse analysis approach (Mobasher et al. 2014a) that uses flexural load-deflection behavior to obtain uniaxial stress-strain behavior. The inverse analysis procedure in described in detail elsewhere (Das et al. 2015a; Mobasher et al. 2014b; a; Soranakom and Mobasher 2008; Soranakom C. and Mobasher B. 2007). The fracture energy and tensile strength values, obtained for different mortars with varying iron powder content, are shown in Table 2. The maximum tensile strength-based criteria is used for damage initiation, and the propagation of crack is modeled using a bilinear traction-separation law (Roesler et al. 2007b) as explained earlier. A mesh-sensitivity study was performed and a mesh containing 1030 CPE4R elements and 1079 nodes provided convergence.

The CDP model (Das and Neithalath 2016; Jankowiak and Lodygowski 2005; Voyiadjis et al. 2008) parameters are $\beta$ or dilatation angle at high confining pressure, $m$ or eccentricity of the plastic potential surface, $\gamma$ that determines the shape of the loading surface in deviatoric plane and s or the ratio of biaxial compressive strength to uniaxial compressive strength of concrete. The values for the above-mentioned parameters are adopted from (Jankowiak and Lodygowski 2005) and shown in Table 2. While these CDP parameters are provided for concrete in (Jankowiak and Lodygowski 2005), they are used in this study for mortars also, for lack of better experimental data on these parameters. The fracture energies and elastic moduli for the different mortars with varying iron powder content is provided in Table 2 while the Poisson's ratio used for simulation is 0.2.

Table 2: CDP and elastic parameters for mortars with iron inclusions

| CDP model parameters [58] | | | | Iron (%) | 0 | 10 | 20 | 30 |
|---|---|---|---|---|---|---|---|---|
| $\beta$ | $m$ | $\gamma$ | $f$ | E (GPa) | 21 | 25.3 | 28.7 | 31.6 |
| 38° | 1 | 0.67 | 1.12 | $G_F$ (N/mm) | 0.033 | 0.034 | 0.038 | 0.045 |

### 3.3.5.3 Numerical Simulation Results:

The XFEM models with the aforementioned inputs were used to yield elastic and fracture outputs. While the load deflection behavior depicts the elastic and post elastic behavior of the mortars with inclusions, the fracture parameters are obtained by their load-CMOD responses, the values of which are further used to calculate the fracture parameters. These fracture parameters when correlated with the experimentally obtained values can provide a clear insight as to the applicability of XFEM model incorporating CDP for studying the influence of cracks and their propagation in quasi-brittle materials like mortars.

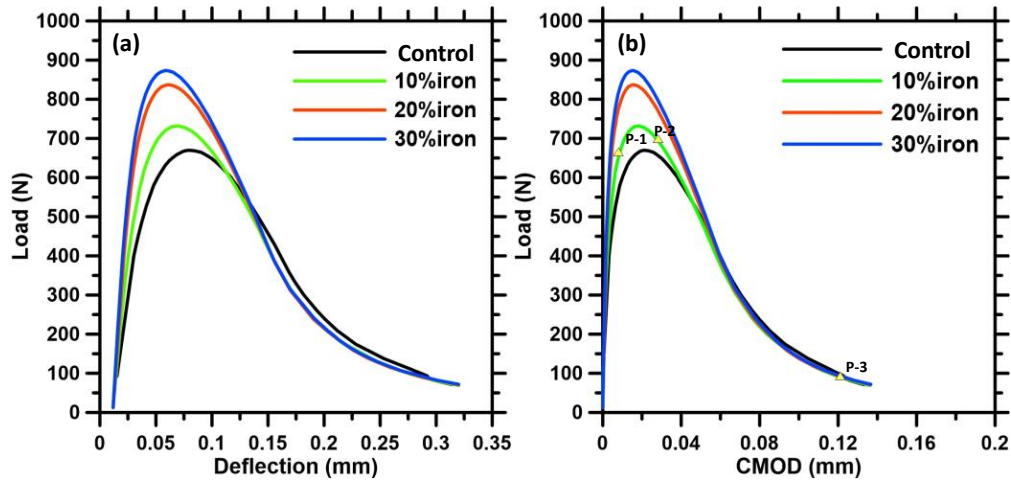

Figure 10: (a) Load deflection and (b) Load CMOD curves indicating P1(pre-peak) at 85% of peak load, P2(post-peak) at 95% of peak load and P3 (near ultimate failure) at 0.125mm CMOD from XFEM model of mortars containing no iron powder, and 10%, 20%, or 30% iron powder

Figure 10 (a) shows the load deflection curves obtained from XFEM analysis. The load-deflection plots show an elastic behavior followed by a non-linear post peak behavior which is a function of the plastic and damage parameters. These graphs depict the predicted behavior for the mortars which, with increasing iron content, show better load carrying capacities, similar to the experimental observations as explained earlier. Figure 10(b) shows the load-CMOD responses obtained from XFEM analysis for control AAS mortar as well as the mortars with varying iron powder contents. As seen in the figure, the load carrying capacity increases with the iron content. The points P-1, P-2 and P-3 for the mortar containing 10% iron powder as slag-replacement correspond to 85% load (pre-peak), 95% load (post-peak) and a load that yields CMOD near ultimate failure respectively. The load-CMOD responses thus obtained, are used later in this section to quantify the fracture responses of these mortars.

Figure 11 shows the maximum principal stress contours (for the mortar containing 10% iron powder) at three different stages of load-CMOD response for the XFEM model incorporating CDP. These stages correspond to three points in Figure 10 (b) where the 85% load is denoted by point P1 (pre-peak), 95% by P2 (post peak) and the point P3 stands for a load that yields a CMOD near ultimate failure. The figure corresponding to point P1 does not show any crack formation since the maximum principal stress failure criterion is not met. On the other hand, the figure corresponding to point P2 show cracks that have grown with a stress concentration at the tip of the crack. The figures corresponding to P3 depict that the cracks have propagated almost completely leading to failure.

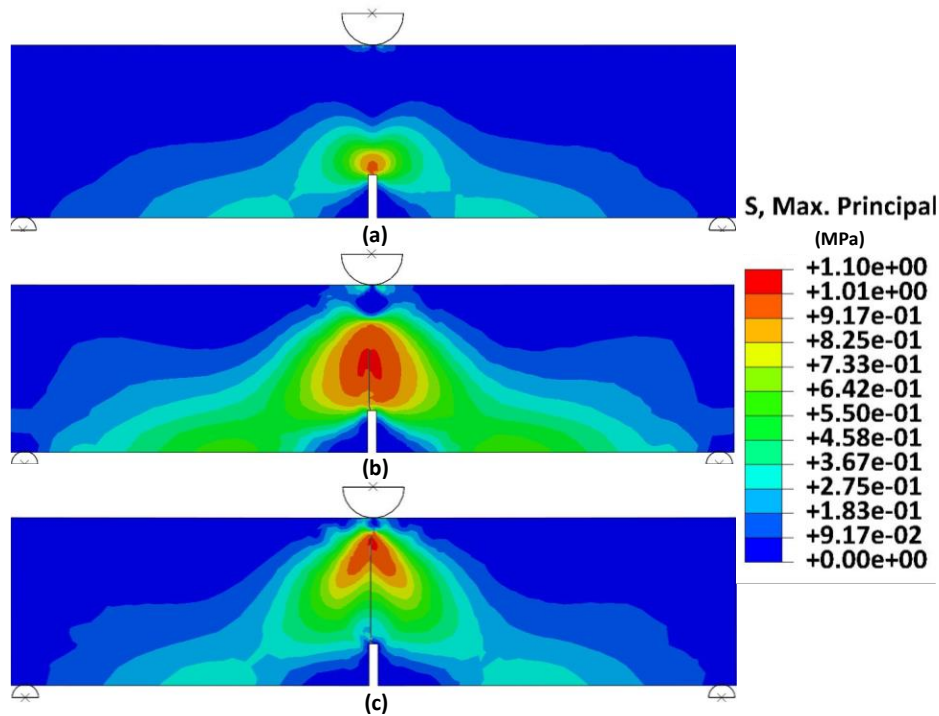

Figure 11: Max principal stress contours for mortar beams with 10% iron (a) P1 (b) P2 (c) P3

To calculate the Mode I critical stress intensity factor $K_{IC}$ (fracture toughness), the crack-extension values, obtained from XFEM as well as DIC (for comparison) at 95% of the peak load in the post-peak regime are used. These can be calculated using the two-parameter fracture model (TPFM) using Equations 13 and 14 (Gdoutos 2006; Nunes and Reis 2012).

$$K_{IC} = \frac{PL}{bd^{3/2}} F[\frac{a_{eff}}{d}] \quad [13]$$

$$F[\frac{a_{eff}}{d}] = [2.9(\frac{a_{eff}}{d})^{1/2} - 4.6(\frac{a_{eff}}{d})^{3/2} + 21.8(\frac{a_{eff}}{d})^{5/2} - 37.6(\frac{a_{eff}}{d})^{7/2} + 38.7(\frac{a_{eff}}{d})^{9/2}] \quad [14]$$

Where the effective crack length $a_{eff} = a_0 + \Delta a$. The CTOD$_c$ or the critical crack tip opening displacement is another fracture parameter from TPFM model that is computed at 95% peak load using XFEM simulation results as well as from DIC as shown in Table 3.

Table 3: Fracture parameters as obtained from the model and DIC

| | Fracture Toughness ($K_{IC}$) [MPa.mm$^{0.5}$] | | | | Crack tip opening displacement, CTOD$_c$ (mm) | | | |
|---|---|---|---|---|---|---|---|---|
| Iron powder % | 0 | 10 | 20 | 30 | 0 | 10 | 20 | 30 |
| XFEM | 20.36 | 21.02 | 22.64 | 28.48 | 0.0078 | 0.0113 | 0.0121 | 0.0142 |
| Experimental (DIC) | 19.50 | 20.26 | 21.58 | 27.28 | 0.0083 | 0.0104 | 0.0106 | 0.0134 |

The fracture parameters ($K_{IC}$ and CTOD$_C$) extracted from the XFEM-based simulation are found to be in very good agreement with those predicted from direct measurements of crack extensions, obtained from DIC for AAS mortar as well as mortars with varying iron powder content. The XFEM-based simulation framework involving maximum principal stress-based crack-initiation criteria along with a crack propagation criteria that implements bilinear traction-separation law in conjunction with the concrete damage plasticity (CDP) model, can be used as an efficient alternative to direct experimental techniques such as Digital Image Correlation (DIC) in these particulate-reinforced AAS mortars. The sensitivity of the CDP model parameters of concrete to the predicted performance of mortars is not very high, as noticed from the predictions.

**4. CONCLUSIONS**

This study shows that elongated metallic iron particulates, generated as a waste product from manufacturing applications can be used as particulate-reinforcement in alkali activated slag mortars with comparable strength and enhanced fracture properties. The elongated iron particles act as micro-reinforcement and improve the crack resistance of the alkali activated slag mortars. The fracture response was characterized through crack growth resistance curves (R-curves) obtained from a compliance-based approach using cyclic crack mouth opening displacement-controlled three-point bending tests on notched mortar beams. The crack growth resistances increased significantly with increasing iron powder content, thereby confirming the beneficial effects of iron powder on the fracture response of alkali activated slag mortars. Alkali-activated slag (AAS) mortars with 30% (by volume) iron powder showed the highest crack-growth resistance. A direct quantification of the fracture process zone parameters (width, length and area of FPZ) was accomplished using Digital Image Correlation (DIC). The significant increase in FPZ area correlated well with the significant increase in crack growth resistances achieved with particulate reinforcement, as increased area of FPZ dissipates significant amount of energy and thus provides improvement in fracture response.

The fracture response of notched beams under three-point bending were simulated using XFEM. The maximum principal stress criteria was adopted in this study for damage initiation. The total fracture energy obtained from a bilinear traction-separation response was used in conjunction with the concrete damage plasticity (CDP) model in ABAQUS$^{TM}$ to simulate crack propagation. The fracture parameters ($K_{IC}$ and $CTOD_C$) extracted from the XFEM-based simulation were found to be in good agreement with those obtained from DIC for AAS mortar as well as mortars with varying iron powder content. The XFEM-based simulation framework provides for direct measurements of crack extensions ($\Delta a$) and fracture responses ($K_{IC}$ and $CTOD_C$) in particulate-reinforced quasi-brittle materials as an efficient alternative to direct experimental techniques such as Digital Image Correlation (DIC).

**ACKNOWLEDGEMENTS**

The authors sincerely acknowledge the support from National Science Foundation (CMMI: 1353170), and the College of Engineering (COE) and Department of Civil and Environmental Engineering at the University of Rhode Island (URI) towards the conduct of this study. The contents of this paper reflect the views of the authors who are responsible for the facts and accuracy of the data presented herein, and do not necessarily reflect the views and policies of NSF, nor do the contents constitute a standard, specification or a regulation. We gratefully acknowledge the use of facilities within the Laboratory for the Science of Sustainable Infrastructural Materials (LS-SIM) and the LeRoy Eyring Center for Solid State Sciences (LE-CSSS) at Arizona State University. Raw materials were provided by Holcim U.S, Schuff Steel, and Iron Shell LLC, which are acknowledged.